\documentclass[figures]{epl}

\usepackage{amsmath}

\newcommand{\be}{\begin{equation}}
\newcommand{\ee}{\end{equation}}
\newcommand{\e}{{\rm e}}

\title{Persistent currents in n--fold twisted Moebius strips}
\shorttitle{Persistent currents in Moebius strips}
\author{E. H. Martins Ferreira\inst{1} \and M. C. Nemes\inst{1} \and M. D. Sampaio\inst{1} \and H. Weidenm{\"u}ller\inst{2}}
\shortauthor{E. H. Martins Ferreira \etal}

\institute{
\inst{1} Department of Physics, ICEx, UFMG - CP 702, 30.161-970, Belo Horizonte, MG, Brazil \\
\inst{2} Max-Planck-Institut f\"ur Kernphysik - D-69029 Heidelberg, Germany
}
\pacs{73.23.–b}{mesoscopic systems}
\pacs{73.23.Ra}{persistent currents}

\begin{document}

\maketitle

\begin{abstract}
We investigate the influence of the topology on generic features of
the persistent current in $n$--fold twisted Moebius strips formed of
quasi one--dimensional mesoscopic rings, both for free electrons and
in the weakly disordered regime. We find that there is no generic
difference between the persistent current for untwisted rings and for
Moebius strips with an arbitrary number of twists.
\end{abstract}

%%%%%%%%%%%%%%%%INTRODUCTION%%%%%%%%%%%%%%%%%%%%%%%%%%%%%%%%%%%
\section{Introduction}

\begin{figure}[p]
\onefigure{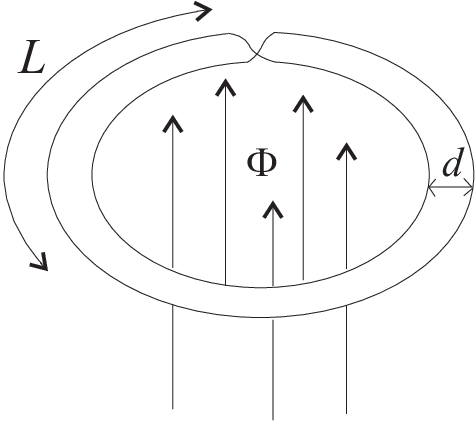}
\caption{Moebius strip of circumference $L$ and width $d$ threaded
  by magnetic flux $\Phi$.} 
\label{fig1}
\end{figure}

Persistent currents in mesoscopic rings threaded by a magnetic flux
offer one of the prime examples of quantum coherence in mesoscopic
physics~\cite{Levy}. The recent experimental realization of rings
which have the shape of singly and doubly twisted Moebius strips and a
diameter of about $50\un{\mu m}$~\cite{Tanda} raises the intriguing
question whether generic features of the persistent current depend
upon the topology of the ring. An affirmative answer would imply a
close connection between topology and quantum coherence. This is the
problem we address in the present paper. We model the ring -- both in
its untwisted and its twisted form -- in the standard way~\cite{Gefen}:
The electrons move either freely or diffusively through a
two--dimensional conducting strip (see fig.~\ref{fig1}). We work in
the regime of zero or weak disorder and assume free motion in the
transverse direction. Our model applies whenever the elastic mean free
path is of the order of or larger than the transverse dimension of the
ring.

Several theoretical papers have dealt with related problems.
Predictions vary according to the dynamical regime under
consideration~\cite{Mila,Hayashi,Deng,Yakubo}. In all these papers,
however, the question whether there is a generic difference
between the untwisted and the twisted case, seems not to have been
addressed. We briefly describe the difference between our model and
that of other recent work at the end of the paper.

We consider a metallic or semiconducting mesoscopic quasi
one--dimensional ring at low temperature. A constant homogeneous
magnetic field $\mathbf{B}$ threads the ring in a direction perpendicular
to the plane of the ring. We take account only of the Aharanov--Bohm
(AB) phase $\phi= 2 \pi \Phi / \Phi_0$. Here $\Phi = AB$ is the
magnetic flux through the ring as given by $B$ and by the area $A$ of
the ring, and $\Phi_0$ is the elementary flux quantum. We disregard
the effect of the magnetic field on the orbital motion of the
electrons in the ring. We also disregard the spin of the electrons.
The ring is twisted $n = 0,1,\ldots$ times. The case $n = 0$
corresponds to the case of an ordinary plane ring, the case $n = 1$ to
an ordinary Moebius strip, and higher values of $n$ to multiply
twisted Moebius strips. We ask: How does the number $n$ of twists
affect the persistent current in the AB ring? 

The case of free electrons and that of diffusive electron motion are
considered in the two next sections. The last section contains a summary 
and a brief comparison with the work of refs.~\cite{Yakubo, Hayashi,Deng}.

\section{Free electrons}
\label{free}

It is useful to investigate first the case of free electrons (no
disorder) at zero temperature, $T = 0$. For simplicity, we take the
ring to be two--dimensional and assume that the transverse extension
$d$ of the ring is very small compared to its circumference $L$. Then,
the ring can be modelled as a rectangle with suitable boundary
conditions at the surfaces. We introduce Cartesian coordinates $x, y$
with $x = [L/(2 \pi)] \theta$ the coordinate in the longitudinal and
$y$ the coordiante in the transverse direction. The valuess $\theta =
0, 2 \pi$ and $y = \pm d/2$ define the surfaces of the rectangle.

The free Hamiltonian has the form
\be
\widehat{H} = -\frac{\hbar^2}{2 \mu}\left[ \frac{(2 \pi)^2}{L^2}
\frac{\partial^2}{\partial \theta^2} + \frac{\partial^2}{\partial y^2}
\right] \ .
\label{1}
\ee
Here $\mu$ stands for the (effective) mass of the electron. The 
single--particle wave functions separate as $\Psi_{j,m}(\theta,y) =
\chi_{j,m}(\theta) \ \psi_j(y)$, and the eigenvalues take the form
\be
E_{j,m} =  E_j + E_j^m \ .
\label{5}
\ee 
The transverse modes $\psi_j(y)$ obey the boundary conditions
\be
\psi_{j}(-d/2) = 0 = \psi_{j}(+d/2) \, .
\label{2}
\ee
The associated eigenvalues $E_j$ are
\be
E_j = \frac{j^2\pi^2\hbar^2}{2\mu d^2} \ .
\label{7}
\ee
The index $j$ defines the channels.

The boundary condition for the longitudinal modes depends on the AB
phase $\phi$. We recall~\cite{Gefen} that for an untwisted ring ($n =
0$), a single traversal of the ring by an electron multiplies the
longitudinal wave function by the AB phase factor $\exp(\pm i\phi)$,
the sign depending upon the direction in which the electron traverses
the ring,
\be
\chi_{j,m}(\theta \pm 2\pi) = \e^{\pm i\phi} \ \chi_{j,m}(\theta) \ .
\label{3}
\ee
In principle, condition~(\ref{3}) applies likewise to twisted rings
($n \neq 0$) but actually becomes modified because we also have to pay
attention to the transverse modes. We demonstrate this for the case $n
= 1$, the ordinary Moebius strip. If the electron starts out at point
$(\theta, y)$, then a single traversal of the ring brings the electron
to the point $(\theta \pm 2 \pi, -y)$. The boundary condition now reads
$\Psi_{j,m}(\theta \pm 2 \pi, -y) = \e^{\pm i \phi} \Psi_{j,m}(\theta,
y)$. The effect of the traversal depends upon whether the transverse
mode $\psi_j$ is even or odd with respect to a reflection of $y$ about
the origin. Even modes (corresponding to odd values of $j$) remain
unchanged while odd modes (even values of $j$) are multiplied by
$(-1)$. Using this fact, we obtain the effective boundary condition
\be
\Psi_{j,m}(\theta \pm 2 \pi, y) = (-1)^{j+1} \e^{\pm i\phi}
\Psi_{j,m}(\theta, y) \ .
\label{4}
\ee
It follows that for a general Moebius strip with $n$ twists, the form
of the eigenfunctions in the longitudinal direction depends only on
whether $n$ is even or odd and not on the actual value of $n$. For $n$
even, the situation is the same as for the case $n = 0$ because an
even number of twists leaves both even and odd transverse
eigenfunctions unchanged. For $n$ odd, the situation is the same as
for the case $n = 1$ because an odd number of twists leaves the even
transverse eigenfunctions unchanged and multiplies the odd ones with
the factor $(-1)$.

\begin{figure}[p]
\onefigure{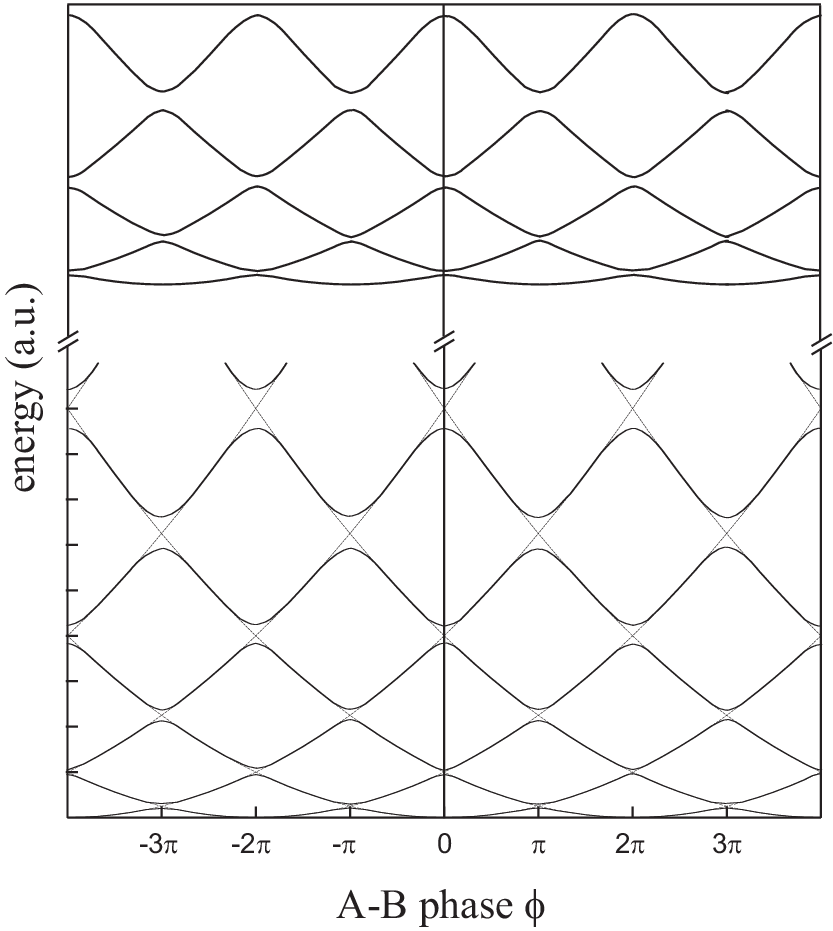}
\caption{Energy as function of the Aharonov-Bohm phase $\phi$ for two
  different transverse modes.} 
\label{fig2}
\end{figure}

Using the boundary condition~(\ref{4}), we obtain for the eigenvalues
$E_j^m$ of the longitudinal modes
\be
E_j^m = \frac{\hbar^2}{2 \mu} \bigg(\frac{2 \pi}{L}\biggr)^2 
\bigg[m + \frac{\phi}{2\pi} + \delta_{j,n} \bigg]^2 \ ,
\label{6}
\ee
with $j$ defined above and $m = 0, \pm 1, \pm 2, \ldots$. The phase
$\delta_{j,n}$ accounts for the symmetry of the transverse modes and
the topology of the ring. It is given by
\be
\delta_{j,n} = \begin{cases}
\frac{1}{2}, & \text{for $j$ even, $n$ odd,} \\
0, & \text{otherwise.}
\end{cases}
\ee
We emphasize the difference between the cases where $n$ is even or
odd. In the first case, the energies in eq.~(\ref{6}), viewed as
functions of $\phi$, form a set of parabolas with minima at $\phi = -
2 \pi m$ with $m = 0, \pm 1, \pm 2, \ldots$. In the second case, we
deal with two classes of parabolas. The energies in channels with odd
values of $j$ form the same kind of parabolas as for even values of
$n$. For channels with even values of $j$, however, the minima of the
parabolas lie at $\phi = - 2 \pi m - \pi$, again with $m = 0, \pm 1,
\pm 2, \ldots$. This situation is illustrated in fig.~\ref{fig2}. It
has to be borne in mind, of course, that the energies $E_j$ for the
even and odd $j$--values are not the same, so that the minima of both
types of parabolas do not have the same values. We conclude that on
the level of the single--particle spectra and their dependence upon
$\phi$, a generic and qualitative difference does exist between
Moebius strips with an even and those with an odd number of twists.
The question is: Does this difference also manifest itself generically
in the persistent current?

At $T = 0$, the persistent current $I(\phi)$ is given by
\be
I(\phi) = -c \frac{\partial E_0}{\partial \Phi} = - \frac{2 \pi
  c}{\Phi_0} \frac{\partial E_0}{\partial \phi} \ .
\label{9}
\ee 
Here $E_0$ is the ground--state energy of a system containing $K$
electrons and is given by
\be
E_0 = \sum_{m,j} p_{j,m} E_{j,m} \ .
\label{8}
\ee
The occupation numbers $p_{j,m}$ are equal to $0$ or $1$, sum up to
$K$, and must be chosen such that $E_0$ is the lowest energy of the
system. From eq.~(\ref{6}) we see that a shift of $\phi \to \phi \pm 2
\pi$ is equivalent to the shift $m \to m \pm 1$. This shows that $E_0$
is periodic in $\phi$ with period $2 \pi$. The persistent current has
that same property. Without loss of generality we may, therefore,
restrict ourselves to the interval $- \pi \leq \phi \leq \pi$.

Before we look at the general case, we first address the situation of
a single channel where the $j$--summation in eq.~(\ref{8}) is
restricted to a fixed value of $j$. We begin with the untwisted case,
$n = 0$. This problem has been thoroughly discussed in
ref.~\cite{Gefen}. Let $K_j$ be the number of electrons in channel $j$.
The form of $E_0$ differs for $K_j = 2 K_0 + 1$ odd and $K_j = 2 K_0$
even. For $K_j$ odd, the sum over $m$ extends from $- K_0$ to $+ K_0$.
The persistent current is accordingly given by 
\be
I(\phi) = - \frac{8 \pi^2 c K_j}{\Phi_0} \frac{h^2}{2 \mu L^2}
\frac{\phi} {2 \pi}\,,\ {\rm for} \ -\pi \leq \phi \leq \pi \ {\rm
  and} \ K_j \ {\rm odd}.
\label{10}
\ee
The persistent current is sawtooth--shaped, with discontinuous jumps
of height $(8 \pi^2 c K_j / \Phi_0) (h^2 / (2 \mu L^2))$ at $\phi =
\pm ( 2 m + 1) \pi$ for all integer $m$. For $K_j$ even, inspection of
eq.~(\ref{6}) shows that the sum over $m$ runs from $- K_0$ to $K_0 -
1$ for $\phi > 0$ and from $- K_0 +1$ to $K_0$ for $\phi < 0$. The
current is
\be
I(\phi) = - \frac{8 \pi^2 c}{\Phi_0} \frac{h^2}{2 \mu L^2} [ K_j
\frac{\phi} {2 \pi} + K_0]\,, \ {\rm for} \ \phi < 0 \ {\rm and} \ K_j \
{\rm even},
\label{11}
\ee
while
\be
I(\phi) = - \frac{8 \pi^2 c}{\Phi_0} \frac{h^2}{2 \mu L^2} [ K_j
\frac{\phi}{2 \pi} - K_0] \,, \ {\rm for} \ \phi > 0 \ {\rm and} \ K_j \
{\rm even}.
\label{12}
\ee
The discontinuity of the sawtooth function is now located at $\phi =
2 \pi m$, with $m = 0, \pm 1, \ldots$. In other words, in comparison
with the case for odd $K_j$, the sawtooth function for even $K_j$ is
shifted by $\pi$. Thus, the current is characteristically different
for even and for odd values of $K_j$ but in both cases is periodic in
$\phi$ with period $2 \pi$.

We keep to the single--channel case and consider next the case of a
singly twisted Moebius strip, $n = 1$. In comparison with the
untwisted case ($n = 0$), there is no change for the ``even'' channels
(odd $j$). However, the situation changes qualitatively for ``odd''
channels (even $j$) because of the changed shape of the spectrum, see
fig.~\ref{fig2}. For $K_j = 2 K_0$ even, the $m$--summation extends
from $- K_0$ to $K_0 - 1$. The current is the same as in
eq.~(\ref{10}). For $K_j = 2 K_0 + 1$ odd, the $m$--summation extends
from $- K_0$ to $+ K_0$ for $\phi < 0$ and from $- K_0 - 1$ to $+ K_0
- 1$ for $\phi > 0$. The current is the same as in eqs.~(\ref{11}) and
(\ref{12}). Thus, for the ``odd'' channels, the persistent current
changes in comparison with the untwisted case: The current for even
$K_j$ in the untwisted case equals the current for odd $K_j$ in the
twisted case, and vice versa. This modification directly reflects the
change in the single--particle spectrum. The generalization to
arbitrary values of $n$ is obvious: Even (odd) values of $n$ produce
the same patterns as $n = 0$ ($n = 1$, respectively). If it were
possible to prepare a Moebius strip with a single ``odd'' channel and
a known number of electrons on it, then the sawtooth pattern of the
persistent current would generically differ for Moebius strips with an
even and an odd number of twists.

Unfortunately, this simple result becomes blurred as we consider the
case of many channels. This is necessary because the number of
electrons on a ring of realistic dimensions is at least several 1000
if not much larger, and the $j$--summation in eq.~(\ref{8}) can,
therefore, not be restricted to a single value of $j$. We first
consider the case of even $n$ and many channels. The number $K_j$ of
electrons in channel $j$ will depend upon the relation between the
energy $E_j$ and the energies $E_k$ of the channels $k \neq j$.
Therefore, the precise form of the current will depend upon both, the
transverse width $d$ of the ring, and the total number $K$ of
electrons on the ring. But the following general statements apply. Let
us first assume that the number of electrons per channel does not
change with $\phi$. This assumption is not realistic but helps to
clarify the situation. Then all channels with an odd number of
electrons will jointly contribute a sawtooth function to the current
which is discontinuous at $\phi = 0$, while all channels with an even
number of electrons will jointly contribute a sawtooth function which
is discontinuous at $\phi = \pm \pi$. The heights of the
discontinuities are given in terms of the total number of electrons in
either case. As a result, the current is periodic in $\phi$ with
period $2 \pi$ but within each period has double--sawtooth structure.
It is only when the numbers of electrons on channels with even $j$ and
with odd $j$ are equal that the period of the current becomes equal to
$\pi$. Our assumption that the number of electrons per channel is
fixed, is unrealistic, however. This is because as we change $\phi$,
there will be crossings of levels pertaining to different channels.
The values of $\phi$ where the crossings occur depend on the
$j$--values of the channels and (through the energies $E_j$) on the
value of $d$ and are, thus, not generic. Such crossings cause the
numbers of electrons per channel to change. As a result, the
persistent current will acquire further sawtooth structures at values
of $\phi$ which differ from both $0$ and $\pm \pi$. However, the
periodicity of the current with period $2 \pi$ will not be affected.
When we average over systems containing different numbers of electrons
(as is done whenever we perfom an ensemble--average over impurities),
the period of the current is expected to be given by $\pi$ as there is
now equal weight given to all possible realizations.

We turn to the case of odd $n$ and many channels. Again, we first
assume that the number of electrons per channel does not change with
$\phi$. In the summation over $j$ in eq.~(\ref{8}), the situation for
channels with odd values of $j$ is the same as described in the
previous paragraph, while the roles of even and odd values of $K_j$
are reversed for even values of $j$. As long as these numbers are
not known individually, the result for the persistent current is
generically the same, however, as for even values of $n$: The
persistent current has sawtooth shape with discontinuities at $\phi =
0$ and at $\phi = \pm \pi$. Further discontinuities arise because the
numbers $K_j$ change with $\phi$. The resulting current is periodic
with period $2 \pi$ and not generically different from that for even
values of $n$.

The effect of finite temperature has been thoroughly discussed in 
ref.~\cite{Gefen}: The occupation probabilities in eq.~(\ref{8}) no
longer are equal to zero or one but are given in terms of Fermi
functions. As a consequence, the sawtooth functions are smoothed, and
the amplitude of the current decreases. But there are no qualitative
changes in the picture that applies for $T = 0$.

\section{Diffusive motion}
\label{diff}

Qualitatively, the influence of diffusion on the results obtained in
the previous section can easily be guessed: Diffusion allows for
elastic scattering between different channels and, thus, removes the
level crossings displayed in fig.~\ref{fig2}. As a result, the
sawtooth pattern of the current (which is a consequence of such level
crossings) is rounded off. The effect increases as the elastic mean
free path decreases. This leads to an ever increasing suppression of
the persistent current. The argument supports the naive expectation
that gross properties of the persistent current like its periodicity
should not depend upon the presence or absence of diffusion. For a
quantitative description, it is necessary to resort to diagrammatic
perturbation theory~\cite{weid1} or to the supersymmetry
technique~\cite{weid2}. We have used the latter method and have
convinced ourselves that the average current is identically the same
for all values of $n$, with $n = 0, 1, \ldots$. The average current is
periodic in $\phi$ as expected. The proof is technically too
complicated to be reproduced here. An extended paper containing the
proof is in preparation~\cite{erlon}.

\section{Conclusions}

We conclude that both for free electrons and for diffusive electron
motion in the weakly disordered regime, there is no generic difference
between the persistent current for untwisted rings and for Moebius
strips with an arbitrary number $n$ of twists. The persistent current
generically has periodicity $2 \pi$ and (at least) a double sawtooth
structure. In detail, the structure depends upon the precise ordering
of the energies of the transverse modes and on the number of electrons
on the ring. For weak disorder, the ensemble--averaged current is
independent of the number of twists.

The work of Yakubo {\it et al.}~\cite{Yakubo} uses a lattice model
with hopping between sites. The authors include both longitudinal and
transverse hopping amplitudes along the ring. We believe that their
numerical work is done in the regime of strong disorder. The authors
conclude that in this regime, the higher Fourier transforms of the
persistent current might allow one to distinguish between a Moebius
strip and an untwisted ring. Transverse hopping is also included in
refs.~\cite{Hayashi,Deng}. The authors conclude that topological
features do affect the physical properties of the system. In both
cases the distinction (if possible) seems to rest upon quantitative
(rather than generic) features of the current. This would be
consistent with our conclusions.

\acknowledgements
This work is supported by Brazilian research agency CNPq and CAPES.
 One of us (HAW) acknowledges support by CAPES
in the form of a CAPES--Humboldt award.

\end{document}